\documentclass[10pt,conference]{IEEEtran}
\usepackage[utf8]{inputenc} 
\usepackage{amsmath}
\usepackage{braket}
\usepackage{graphicx}
\usepackage{pgfplots}
\pgfplotsset{compat=1.15}
\usepackage{csquotes}
\usepackage{hhline}
\usepackage{amssymb}

\usepackage{dcolumn}
\usepackage{tabularx}
\usepackage[colorlinks=true,linkcolor=blue,citecolor=blue,urlcolor=blue]{hyperref}
\usepackage{braket}

\usepackage{xparse}
\NewDocumentCommand{\tens}{e{_^}}{%
  \mathbin{\mathop{\otimes}\displaylimits
    \IfValueT{#1}{_{#1}}
    \IfValueT{#2}{^{#2}}
  }%
}
\usepackage{listings}
\usepackage{color}
\usepackage{subcaption}
\usepackage{placeins}
\definecolor{mygreen}{rgb}{0,0.6,0}
\definecolor{mygray}{rgb}{0.5,0.5,0.5}
\definecolor{mymauve}{rgb}{0.58,0,0.82}
\usepackage{graphicx}% Include figure files
\usepackage{dcolumn}% Align table columns on decimal point
\usepackage{bm}% bold math
\usepackage{tikz-cd}
\usetikzlibrary{arrows}
\usepackage{tikz}
\usepackage{pgfplots}
\usepackage{lipsum, babel}
\pgfplotsset{compat=newest}
\usetikzlibrary{decorations.markings}
\usepackage{pst-solides3d}
\usepackage[utf8]{inputenc}
\usepackage{graphicx}
\usepackage{braket}
\usepackage{dsfont}
\usepackage{amsmath} 
\usepackage[numbers,sort&compress]{natbib} 
\usetikzlibrary{shapes.geometric}
\usepackage[utf8]{inputenc}

\usepackage{tikz-cd}
\usetikzlibrary{arrows}
\usepackage{tikz}
\usepackage{pgfplots}
\usepackage[utf8]{inputenc}
\usepackage{tikz}
\usepackage{qcircuit}
\usepackage{caption}
\usetikzlibrary{arrows.meta,
                chains,
                positioning,
                shapes.geometric}
\usepackage{siunitx}
\usepackage{mathtools}

\tikzset{FlowChart/.style={
startstop/.style = {rectangle, draw, fill=red!30,
                    minimum width=1cm, minimum height=1cm,
                    on chain, join=by arrow},
  process/.style = {rectangle, rounded corners, draw, fill=blue!30,
                    text width=5cm,  minimum height=1cm, align=center,
                    on chain, join=by arrow},
    arrow/.style = {thick,-Triangle}
        }   }
\pgfplotsset{compat=newest}
\usetikzlibrary{decorations.markings}
\usepackage{pst-solides3d}
\usepackage[utf8]{inputenc}
\usepackage{graphicx}
\usepackage{braket}
\usepackage{dsfont}
\usepackage{amsmath} 
\usepackage{natbib} 
\usetikzlibrary{shapes.geometric}
\usepackage[utf8]{inputenc}

\tikzstyle{startstop} = [rectangle, rounded corners, minimum width=3cm, minimum height=0.5cm,text centered, draw=black, fill=red!30]
\tikzstyle{io} = [trapezium, trapezium left angle=70, trapezium right angle=110, minimum width=3cm, minimum height=0.5cm, text centered, draw=black, fill=blue!30]
\tikzstyle{process} = [rectangle, minimum width=3cm, minimum height=0.5cm, text centered, draw=black, fill=orange!30]
\tikzstyle{decision} = [diamond, minimum width=1.5cm, minimum height=0.3cm, text centered, draw=black, fill=green!30]
\tikzstyle{arrow} = [thick,->,>=stealth]
\tikzstyle{rec}=[to path={(\tikztostart) -- ++(#1,0pt) \tikztonodes |- (\tikztotarget) },pos=0.5,->,>=stealth, thick]
\tikzstyle{descr}=[inner sep=2.5pt]
\newcolumntype{C}{>{\centering\arraybackslash}X}

\begin{document}
\title{Fully Quantum Hash Function}
\author{
    \IEEEauthorblockN{Shreya Banerjee\IEEEauthorrefmark{1}, Harshita Meena\IEEEauthorrefmark{2}, Somanath Tripathy\IEEEauthorrefmark{2}, Prasanta K. Panigrahi\IEEEauthorrefmark{1}\IEEEauthorrefmark{3}}
    \IEEEauthorblockA{\IEEEauthorrefmark{1} Center for Quantum Science and Technology\\
    Siksha 'O' Anusandhan University, Bhubaneswar-751030, Odisha, India
    }
    \IEEEauthorblockA{\IEEEauthorrefmark{2}Department of Computer Science and Engineering, Indian Institute of Technology Patna-801106, Bihar, India
    }

    \IEEEauthorblockA{\IEEEauthorrefmark{3}Department of Physical Sciences,\\ Indian Institute of Science Education and Research Kolkata,\\ Mohanpur-741246, West Bengal, India
   }
}

\maketitle

\begin{abstract}
We introduce a novel, \textit{fully} quantum hash (FQH) function  within the quantum walk on a cycle framework. We incorporate the deterministic quantum computation with single qubit to replace classical post-processing, thus increasing the inherent security. Further, our proposed hash function exhibits zero collision rate and high reliability. We further show that it provides $ > 50\%$ avalanche in average, and is highly sensitive to the initial conditions. We show comparisons of several performance metrics for the proposed FQH on different settings as well as with existing protocols to prove its efficacy. FQH requires minimal quantum resources to produce a large hash value, providing security against the birthday attack. This innovative approach thus serves as an efficient hash function and lays the foundation for potential advancements in quantum cryptography by integrating the fully quantum hash generation protocol.
\end{abstract}

\IEEEpeerreviewmaketitle

\section{Introduction}
As the digital landscape evolves, the need for robust cryptographic mechanisms to safeguard sensitive information increases with it. In the 1950s, Knuth \cite{b1} first introduced the concept of hash functions. A hash function is a mathematical function that takes a string of bits as input ('message') and returns a fixed-size string of bytes ('hash'). The output, typically a hash value or hash code, is a compact input data representation. Hash functions are designed to be fast to compute and to produce a unique hash value for each unique input. The key properties of hash functions include:  deterministic generation, fast computation, fixed output size, pre-image resistance, collision resistance etc. Whereas modern cryptographic hash functions such as SHA-256 (Secure Hash Algorithm 256-bit) or MD5 (Message Digest Algorithm 5) are preferred \cite{b2} for practical implementations;  several classical hash functions, including MD5, SHA-1, and RIPEMD, have been found to have significant security vulnerabilities related to collision frequencies \cite{b3,b4,b5}. Moreover, the advent of quantum computers poses a formidable challenge to the security infrastructure that relies on classical cryptographic primitives. Though reliable in the classical computing realm, classical hash functions are susceptible to swift decryption by quantum algorithms, threatening the essence of data integrity and security \cite{blckchain, Xu2021, Selvarajan2023}.
In cryptography and data integrity verification, hashing algorithms play a pivotal role in ensuring the security and authenticity of digital information \cite{b6,b7,b8}. Traditional hashing techniques, while effective, are often computationally intensive and vulnerable to attacks from increasingly sophisticated adversaries. Quantum computing, with its promise of exponential processing power, presents a tantalizing avenue for redefining cryptographic primitives. We can easily exploit these classical hash functions using the powerful parallelism capability of quantum computation, such as Shor’s algorithm for factoring and Grover’s algorithm for database search \cite{b9,b10}. Hence, quantum cryptography emerges as a promising solution to overcome these vulnerabilities \cite{b11,b12}.
Quantum hash functions (QHFs) represent a crucial aspect of quantum cryptography, garnering significant attention \cite{b13,b14,b15,b16,b17,b18,b19}. QHFs transform classical messages into a Hilbert space, aiming to minimize the space to prevent excessive information leakage, as stipulated by quantum principles like Holevo-Nayak’s theorem and the Holevo limit \cite{nayak1999optimallowerboundsquantum}. Additionally, QHFs endeavor to maximize the quantum distance between images of distinct classical messages to prevent collisions. Originally introduced implicitly in Buhrman et al. \cite{b13} as quantum fingerprinting employing binary error-correcting codes, subsequent work by Gavinsky et al. \cite{b14} recognized the poterntial of QHF potential as a cryptographic primitive. Ablayev et al. defined and constructed non-binary QHFs \cite{b15} and introduced a balanced variant \cite{b16}, albeit dependent on quantum entanglement. Ziatdinov \cite{b17} extended quantum hashing to arbitrary finite groups, presenting novel constructions based on expander graphs and extractor functions \cite{b18}. Existing QHFs with classical input and quantum hash value rely on the SWAP-test for verifying equality between hash values, with a one-sided error probability of 0.75 \cite{b19}. Achieving arbitrarily low error probabilities necessitates encoding n-bit classical messages into $\log(\log(1/\epsilon))$ qubits, where $\epsilon > 0$ \cite{b19}.

Quantum walks (quantum walks) have garnered considerable attention owing to their intrinsic ability to faster mixing and wide-ranging potential applications, with experimental demonstrations already substantiating their viability \cite{b20,b21,b22,b23}. Li et al.~\cite{b24} pioneered a quantum hash function (QHF) grounded in two-particle interacting quantum walks, incorporating $I$ and $\pi$-phase interactions guided by message bits. Building upon this foundation, our study presents a novel and innovative approach termed 'Fully Quantum Hash Function (FQHF)', where we leverage the discrete-time quantum walk on cycles, and dense coding of coin operators, similar to~\cite{b24, b25, b26}. However, these previous approaches relies heavily on classical post-processing to obtain the hash value, which leaves the schemes vulnerable to quantum attacks during these steps. In this work, we incorporate a scheme based on the deterministic quantum computation with (one) clean qubit (DQC1)~\cite{DQC1} into this framework as our novel contribution. We show that the inclusion of DQC1 in quantum hash generation, it is possible to replace the classical post-processing with a prepare and measure protocol, as DQC1 computes the expected outcome of random unitary matrices with respect to a given quantum state. We further show that the generated hash performs extremely well under a wide range of tests for careful selection of random unitary and system parameters. 

The paper is organized as follows. In section~\ref{Prelims}, we have provided necessary prerequites regarding discrete time quantum walk, DQC1, and a breif discussion on the random unitary ensembles used in this work. In section~\ref{SFQHF}, we present our Hash generation protocol with example. In section~\ref{RnD}, we discuss our results and performance of our generated hash, subject to different tests. We finally conclude in section~\ref{conc} with future directions. 

\section{Prerequisites}\label{Prelims}
\subsection{Discrete-time quantum walk.} 
The time evolution of the discrete-time quantum walk on a line is controlled by the random flip of a coin, and a shift operator that is controlled by the state of the coin, i.e., the walker takes a step in the forward direction if the outcome of the coin is \textit{head}, and backward direction if the outcome of the coin is \textit{tail}. Thus the shift operator is comprised of two different operators, \textit{increment} and \textit{decrement}. 
\begin{figure} [hbt!]
    \centering
    \includegraphics[width=0.7\linewidth]{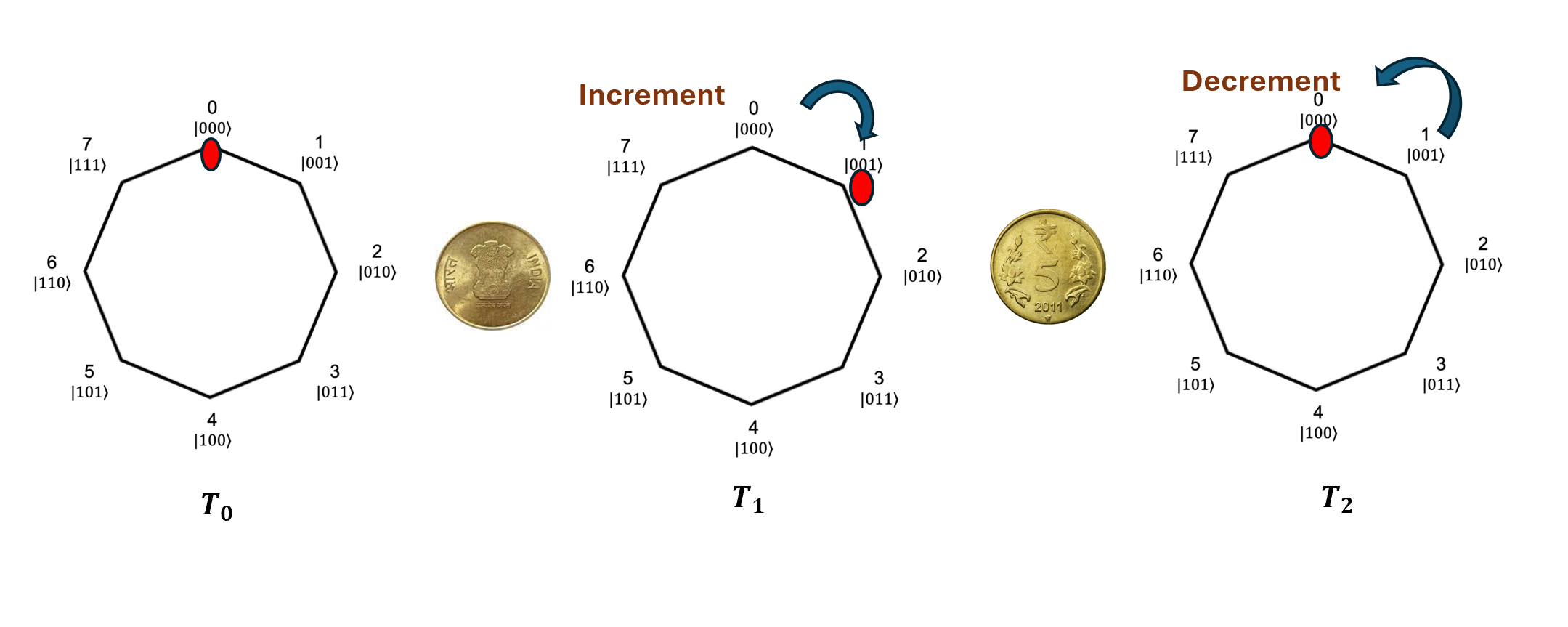}
    \caption{Schematic of discrete time quantum walk on a cycle with $8$ nodes. $T_0$, $T_1$, and $T_2$ represents the time. If the coin output is head, the walker takes a step in the clockwise direction (increment), and if it is tail, the walker takes a step in the anticlockwise direction (decrement).}
    \label{qwc}
\end{figure}

\begin{figure}[hbt!]
    \centering
    \includegraphics[width=0.8\linewidth]{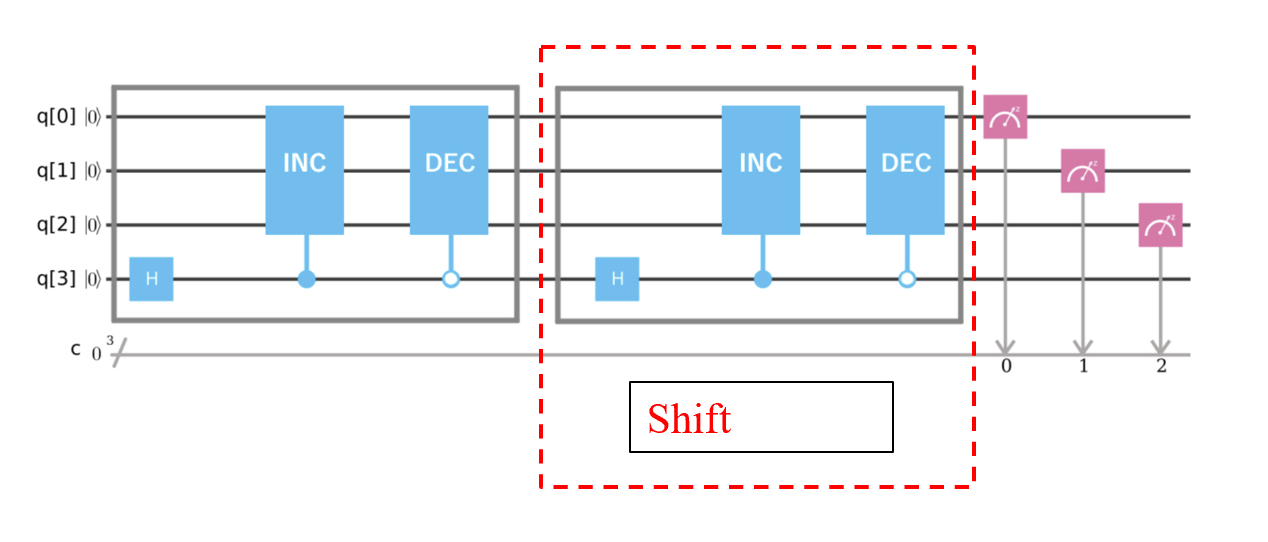}
    \caption{Quantum circuit to realize the quantum walk in Fig.~\ref{qwc}. Three qubits ($q[0], q[1], \text{and} q[2]$) encodes the $8$ position nodes; $q[3]$ encodes the coin. Here Hadamard coin is used for the walk.} 
    \label{Cqwc}
\end{figure}

For one-dimensional discrete-time quantum walk on a line, the most common coin operator can be written as
\[
C = \frac{1}{\sqrt{2}} \begin{bmatrix}
    1 & 1 \\
    1 & -1
\end{bmatrix}, \tag{1}
\]
which is the unbiased Hadamard operator. The conditional shift operator $S$ is given by,
\[
\hat{S} = \sum_x \left(\langle x+1, 0| \, |x, 0\rangle + \langle x-1, 1| \, |x, 1\rangle \right) \tag{2};
\]
here the sum is taken over all positions $x$ that can be taken by the walker after taking each step. Thus, the conditional shift operator $S$, decides the direction in which the walker moves.
The one-dimensional discrete quantum walk on a cycle with $N$ position nodes uses the conditional shift operator $S_c$ given as, 
\begin{align*}
S_c &= \sum_{x=1}^{N-1} \left(\langle x+1, 0| \, |x, 0\rangle + \langle 1, 0| \, |N, 0\rangle \right) \\
&\quad + \sum_{x=2}^{N} \left(\langle x-1, 1| \, |x, 1\rangle  + \langle N, 1| \, |1, 1\rangle \right). \tag{3}
\end{align*}
Here, the increment can be though of as a step in the clockwise, and decrement can be thought of as a step in the anticlockwise direction. The simplistic quantum walk on a cycle with $8$ nodes and its quantum circuit is shown is Fig.~\ref{qwc} and Fig.~\ref{Cqwc} respectively. Quantum random walk on a graph has the possibility to provide an advantage over its classical counterpart as a quantum walker walks faster on certain graphs \cite{Melnikov_2019}. For a cycle, quantum walk has a quadratic speedup with respect to mixing time \cite{Childs2002}.

\subsection{Dense coding of coin-operators for input message encryption.}
An efficient quantum hash function can be constructed by subtly modifying discrete-time quantum walk on a cycle \cite{b24,b25,b26}. Usually, the coin operator at each step of the walk is made dependent on a binary string of a user-fixed length. Considering the length of string be $2$, four different coin operators $C_1$, $C_2$, $C_3$, and $C_4$ can be used which are controlled by the message bits $"00", "01", "10",$ and $"11"$ respectively. We use a rotation operator with $4$ different angle to encode the coin operators in this work, i.e., 
\[
C_i= 
\begin{bmatrix}
    \cos\theta_i & \sin\theta_i \\
    \sin\theta_i & -\cos\theta_i
\end{bmatrix},
\]
 where, $i \in \{00, 01, 10, 11\}$. As an example, if the input message is "10001101", the walker will take $4$ steps, and the coin operators $C_{10}$, $C_{00}$, $C_{11}$, and $C_{01}$ will be applied on the coin register sequentially. 
\subsection{Deterministic Quantum Computation with one qubit (DQC1).}
\begin{figure}[h!]
    \includegraphics[scale=0.4]{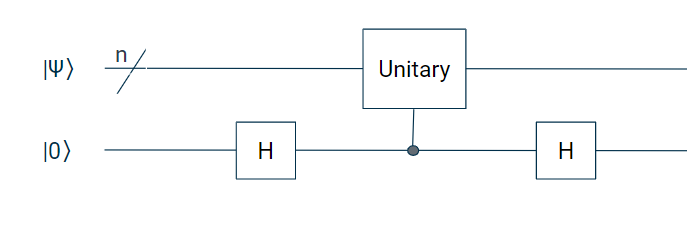}
    \caption{
    Circuit diagram for DQC1.}
    \label{Htest}
\end{figure}

The core principle of the deterministic quantum computation with one clean qubit \cite{DQC1} rests on the ability to translate the expectation value calculation of a unitary operator $U$ with respect to a $n-$qubit state $\ket{\psi}$ into the probability of a certain measurement outcome \cite{b27,b28,b29,b30}. It can be realized in the following steps.
\begin{itemize}
    \item \textbf{Initialization:} An ancilla qubit is prepared in the $|0\rangle$ state.
    
    \item \textbf{Superposition:} A Hadamard gate ($H$) is applied to the ancilla qubit, creating a superposition: 
    $$\frac{1}{\sqrt{2}}(|0\rangle + |1\rangle) \otimes |\psi\rangle$$.
    
    \item \textbf{Controlled-U:} The unitary operation $U$ is applied to the state register as shown in Fig.~\ref{Htest} with the ancilla qubit as a control. This entangles the ancilla qubit with the system. The state of the conjugate system is now, 
    $$\frac{1}{\sqrt{2}}(|0\rangle \otimes |\psi\rangle + |1\rangle \otimes U|\psi\rangle)$$.
    
    \item \textbf{Second Hadamard:} Another Hadamard gate is now applied to the ancilla qubit.
    
    \item \textbf{Measurement:} The single qubit ancilla register is now measured in the computational basis ($\{|0\rangle, |1\rangle\}$).
\end{itemize}

The corresponding circuit diagram is provided in Fig.~\ref{Htest}. 

The probability of measuring the ancilla qubit in the $|0\rangle$ state is given by, 
$P(|0\rangle) = \frac{1}{2} \left(1 + \text{Re}(\langle \psi | U | \psi \rangle)\right).$ As one can see, the expectation value $\text{Re}(\langle \psi | U | \psi \rangle)$ (the real part) can be estimated with a single qubit through DQC1. To obtain the imaginary part, $\text{Im}(\langle \psi | U | \psi \rangle)$, a slight modification to the circuit is needed, i.e., one needs to apply an $S^\dagger$ gate (phase gate) after the initial Hadamard gate. Through this scheme, the probability of measuring $\ket{0}$ at the single qubit ancilla encapsulates the entire expectation value of an $n-$qubit operator with respect to a $n-$qubit state.  

\subsection{Random Unitary Matrices}
Random unitary matrices are fundamental in various quantum algorithms and protocols. In various protocols, the choice of random unitary matrices ensures effective entanglement within the system, as well as help increasing security of a cryptographic protocol through their inherent randomness. In this work, we have used two different types of widely used random unitary matrices from the circular ensemble, mainly due to their polynomial implementability through unitary-T-design \cite{Haferkamp2022randomquantum, Harrow2023}. In the context of random matrix theory, the circular ensembles are measures on spaces of unitary matrices introduced by Freeman Dyson \cite{DysonF}. The circular unitary ensemble (CUE) and the circular orthogonal matrices are the Haar measures on unitary matrices and symmetric unitary matrices respectively \cite{Haar}. Each ensemble plays a crucial role in various areas of physics, including quantum chaos, disordered systems, and quantum information theory \cite{b31,b32,b33}.

\section{Fully Quantum Hash Generation Protocol}\label{SFQHF}
We briefly describe the fully quantum Hash generation protocol. Assuming a $2l-$bit message as, \( M = (m_1 m_2\ldots m_{2l}) \), we explain the procedure of generating a hash value from it.

\begin{figure}[hbt!]
   \includegraphics[scale=0.15]{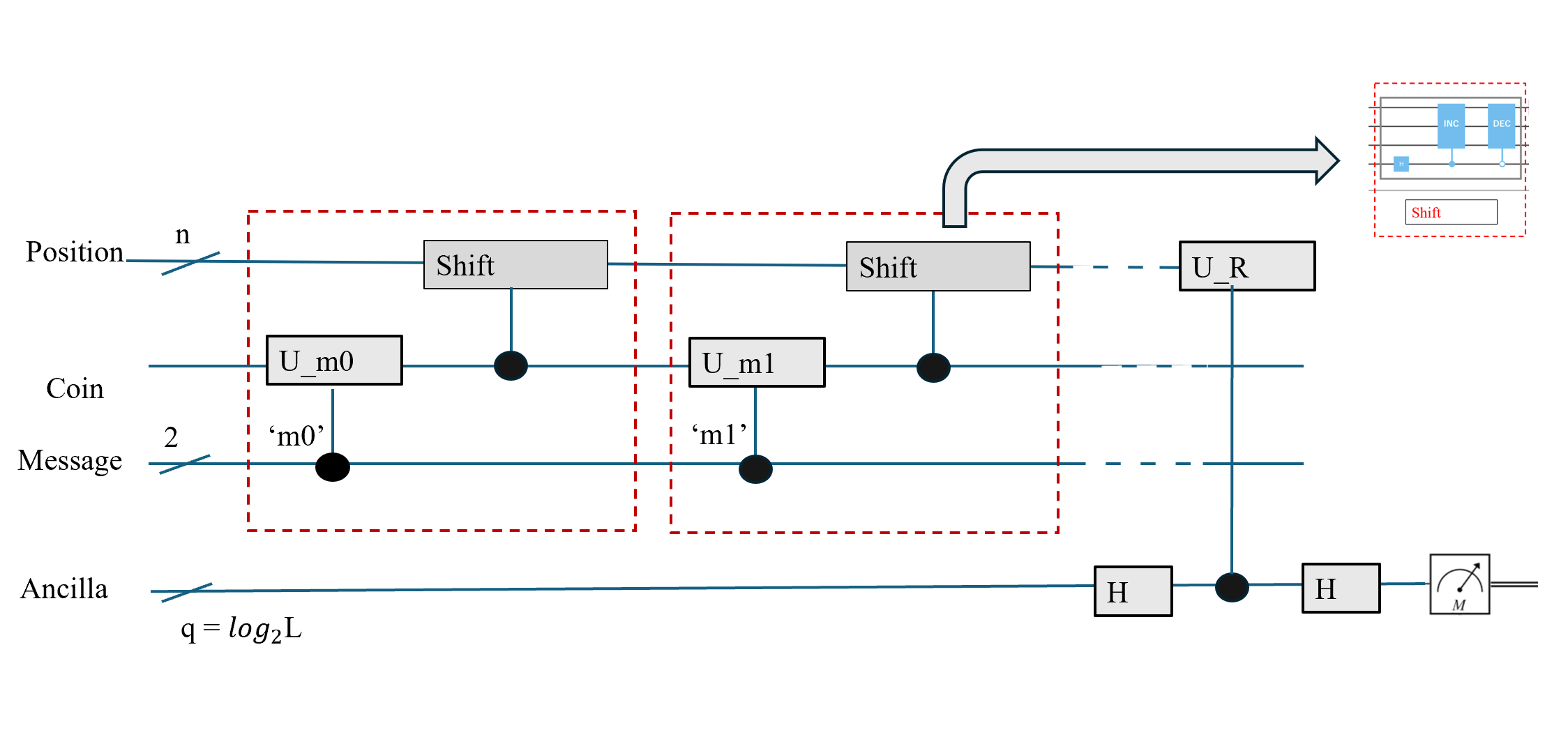}
    \caption{Schematic circuit for Fully Quantum Hash Generation. 'm0' and 'm1' refers to the first and second pair of bits in the message. The Shift operator encodes one single step of the quantum walker. $L$ is the length of generated Hash upto a scalar factor. $U_R$ refers to the random unitary matrix applied on the position register with a control on the ancilla register.}
    \label{FQHF}
\end{figure}

\begin{figure}[hbt!]
   \includegraphics[scale=0.12]{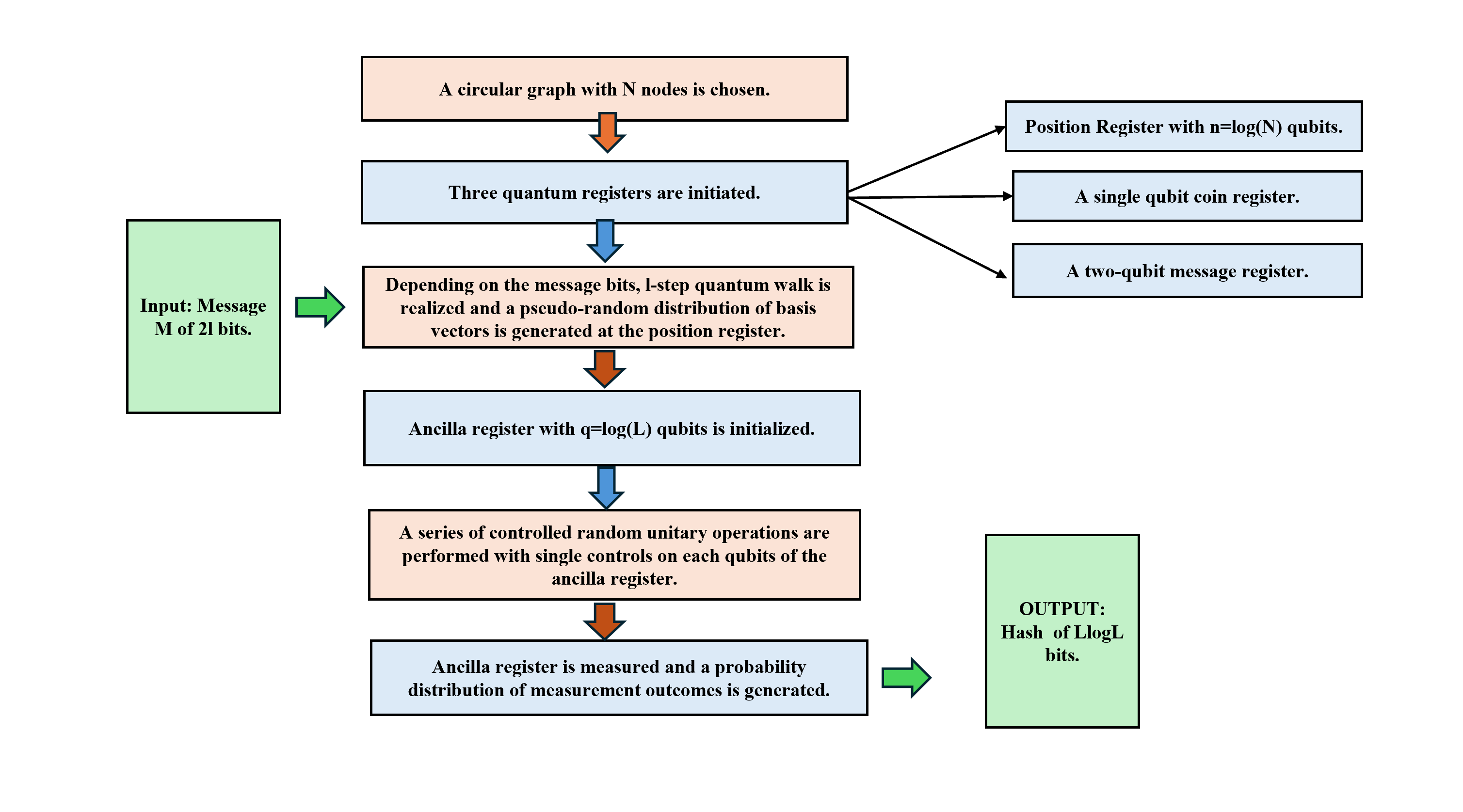}
    \caption{Flowchart for Fully Quantum Hash Generation.}
    \label{flow}
\end{figure}

 The protocol uses an input message with an even number of bits, and if the length of the message is not an even number, a single bit "0" is added at the end of the message. We use three different quantum registers for the protocol: a position register to generate a quantum state constituting basis vectors with random amplitudes, a single qubit quantum coin register, and an ancilla register to implement the series of Hadamard tests. A thematic circuit diagram for the protocol is provided in Fig.~\ref{FQHF}.

\begin{itemize}
    \item[1.] Our protocol uses a simple discrete-time quantum walk on a cycle as the preliminary random state generator. Based on the number of nodes on the cycle $N$, the number of qubits in the position register $n$ is chosen to be $n=\log(N)$. Consequently, we choose to associate two bits of the input message with different steps of the walker and represent them with $q$ randomly selected different angles, i.e., $\theta_1, \theta_2,...,\theta_q$, where $0 < \theta_i < \frac{\pi}{2} \forall i$.
    
    \item[2.] Next, the one-coin, one-walker, discrete-time quantum walk on a cycle with $n$ nodes under the control of the message $M$ (Fig.~ \ref{FQHF}) is executed, as explained in Sec.~\ref{Prelims} and a quantum superposition of basis vectors with a pseudo random distribution is generated at the position register.
    
    \item[3.] Consequently, we implement a series of controlled random unitary operations on the qubits of the position register, with the controls on individual qubits in the ancilla resister. The nature of target unitary operations can be diverse, i.e., one can use random unitaries of different size and types as the control to achieve different outcomes. We provide an analysis on the security of Hash Function based on different types and sizes of random unitaries in Sec.~\ref{RnD}.
    
    \item[4.] Post implementation of the random controlled unitaries, we measure the ancilla register multiple times to obtain a probability distribution of constituent basis states. Further, we sort them based on the descending order of probability, and concatenate the basis states to obtain the final Hash function. As, each basis state consists of $2^q$ bits, where $q$ is the number of qubits present in the ancilla register, the length of the hash generated through this process is $q\times 2^q$. A discussion on the required number of ancilla register measurement is done in the Sec.~\ref{RnD}. 
\end{itemize}

Fig.~\ref{flow} provides a schematic diagram of the workflow of the protocol. 

\section{Results and Discussion}\label{RnD}
We consider standard metrics to analyse the performance of our protocol, for two different lengths of the output hash value, 160-bit and 384-bit. Following our protocol, the spatial costs to generate a 160-bit Hash and a 384-bit Hash are $5 + 1 + 2 + 5=13$ and $5 + 1 + 2 + 6=14$ qubits from an input message of arbitrary length, where there are $5$ position qubits, $1$ coin qubit, and $2$ message qubits in both cases, and $5$, and $6$ are the number of qubits in the ancilla register respectively. We have used the statevector simulator with Qiskit for a proof of concept, noiseless simulation of our theoretical protocol.  

\subsection{160-bit Hash Generation and Analysis:}\label{H160}
We first present our results for the 160-bit Hash generation.

\subsubsection{Sensitivity of Hash value to Message}
The sensitivity of a hash function refers to how much the output hash value changes when there are slight variations in the input message. A good hash function should exhibit high sensitivity, meaning that even small changes in the input message should result in significantly different hash values. This property is essential for ensuring the security and reliability of the hash function, as it prevents adversaries from predicting or manipulating the hash output.
To assess the sensitivity of the hash value to variations in the message, we consider the following conditions:

\begin{enumerate}
    \item \textbf{Condition 1:} The original message.
    \item \textbf{Condition 2:} Change a bit of the original message from “0” to “1” at a random position.
    \item \textbf{Condition 3:} Change a bit of the original message from “1” to “0” at a random position.
    \item \textbf{Condition 4:} Delete the first bit of the message.
    \item \textbf{Condition 5:} Insert a bit into the original message randomly.
\end{enumerate}

\begin{table}[hbt!]
    \centering
    \begin{tabular}{c|c}
    \hline \\
        \textbf{Condition} & \textbf{Generated Hash when CUE is used} \\
    \hline \\
        1 &  '082292C10450963766AFC368CD4B6699FDFEF2D7'\\
    \hline \\
        2 & '2A88B10D28081B563B547EE5331F388C3BCF7ED7' \\
    \hline \\
        3 & '8230290CA3ADA260EA97510FB6AFBECB93CFB1F' 
     \\
     \hline \\
        4 & 'CE2D7946704F51CAD0C7F0BE11834DDB0A473D4B' \\
    \hline \\
         5 &'4A0626E598098E09C9CF52FBC8DAF0293DFD6EB4' \\
    \hline 
    \end{tabular}
    \caption{Output hash in Hexadecimal format for input message $'01100010110101001'$ for different conditions, generated with 5 position qubits, 1 coin qubits, and 5 ancilla qubits. The controlled unitary matrices after quantum walk are $5$ random $2 \times 2$ matrices from Circular Unitary Ensemble with a control on each ancilla qubit, applied to random qubits on the position register.}
    \label{Sensitivity_CUE}
\end{table}

\begin{table}[htb!]
    \centering
    \begin{tabular}{c|c}
    \hline \\
        \textbf{Condition} & \textbf{Generated Hash when COE is used}  \\
    \hline \\
        1 & '4331C308047ACE353AD2853FBBCF5E6A4A1EE6B1'\\
    \hline \\
        2 & '4300430B1C38DEB852D2BCD4EFEDA9D78A1EE6B1' \\ 
    \hline \\
        3 & '430C2C7004B4A14538677AF5E9DFFB4B425CF635'\\
    \hline \\
        4 & '010E330A14BCED27AD0C87FBC712DAC54ECF75E'\\
    \hline \\
        5 & 'FEDEBC710CEE677A41A91E8FE20152B3A3511825'\\
    \hline
    \end{tabular}
    \caption{Output hash in Hexadecimal format for input message $'01100010110101001'$ for different conditions, generated with 5 position qubits, 1 coin qubits, and 5 ancilla qubits. The controlled unitary matrices after the quantum walk are $5$ random $2 \times 2$ matrices from Circular Orthogonal Ensemble with a control on each ancilla qubit, applied to random qubits on the position register.}
    \label{Sensitivity_COE}
\end{table}
We present our outcome hash in hexadecimal format for a single input message $'01100010110101001'$. We have considered consider two different types of random unitary matrices as the controlled unitary applied on the pseudo-random superposed quantum state. The obtained hash using random CUE and COE matrices of dimension $2 \times 2$ as targets of the controlled unitary part of the protocol under each of the aforementioned conditions is provided in Tables~\ref{Sensitivity_CUE} and \ref{Sensitivity_COE}. As can be seen from the tables, our protocol exhibits excellent sensitivity for all five conditions in each of the matrices. We investigate the collision resistance property to have ensure the security of the Hash generation protocol. 

\subsubsection{Collision Resistance}

\begin{table}[htbp]
    \centering
    \begin{tabular}{ccccc}
    \hline \\
        \textbf{Matrix}& \textbf{Dimension} & \textbf{Qpos} & \textbf{Total Collision} & \textbf{Collision Rate} \\
    \hline \\
        CUE & $2 \times 2$ & 5 & 0 & 0\\
        CUE &$2 \times 2$ & 6 & 0 & 0 \\
        CUE &$2 \times 2$ & 7 & 0 & 0 \\
        CUE &$4 \times 4$ & 5 & 0 & 0 \\
        CUE &$4 \times 4$ & 6 & 0 & 0 \\
        CUE &$4 \times 4$ & 7 & 0 & 0 \\
        CUE &$8 \times 8$ & 5 & 0 & 0 \\
        CUE &$8 \times 8$ & 6 & 0 & 0 \\
        CUE &$8 \times 8$ & 7 & 0 & 0 \\
        \hline \\
        COE & $2 \times 2$ & 5 & 0 & 0\\
        COE & $2 \times 2$ & 6 & 0 & 0 \\
        COE & $2 \times 2$ & 7 & 0 & 0 \\
        COE & $4 \times 4$ & 5 & 0 & 0 \\
        COE & $4 \times 4$ & 6 & 0 & 0 \\
        COE & $4 \times 4$ & 7 & 0 & 0 \\
        COE & $8 \times 8$ & 5 & 0 & 0 \\
        COE & $8 \times 8$ & 6 & 0 & 0 \\
        COE & $8 \times 8$ & 7 & 0 & 0 \\
        \hline\\
        \end{tabular}
        \caption{Collision Analysis table for  of Different Input types. Total number of times Hash is generated for each individual case (each row of the table) is $1024$.}
        \label{collision}
\end{table}
Collision resistance is a property of hash functions that ensures it is computationally infeasible to find two different inputs that creates the same output hash. In other words, a hash function is considered collision-resistant if it is difficult to find two distinct inputs $x$ and $y$ such that $H(x) = H(y)$, where $H$ is the hash function. To establish the said infeasibility of computation, we follow the procedure given below. 
\begin{itemize}
    \item[1.] Select a message randomly and generate the corresponding hash value.
    \item[2.] Change the message randomly by flipping a bit and generating the corresponding hash value.
    \item[3.] Compare these two hash values to check if they are equal.
    \item[4.] If the two hash values are equal, increase the total collision count by 1.
    \item[5.] Repeat steps 1 to 4 $1000$ times.
    \item[6.] Collision Rate (CR) is computed as, \begin{equation}
        \text{CR} = \frac{\text{Total Collision count}}{1000}.
    \end{equation} 
\end{itemize}
We have performed the test considering different aspects of our Hash generation protocol, i.e., changing type and dimension of the target unitary matrix assumed, and with varying positional qubits. Results are provided in Table \ref{collision}. As can be seen from Table~\ref{collision}, We found no collision in any of the cases we investigated. 

\subsubsection{Avalanche effect}
The avalanche effect is a desirable property of cryptographic hash functions and other cryptographic algorithms. It refers to the significant and unpredictable change in the output (hash value) when even a tiny change is made to the input. In other words, a minor alteration in the input should lead to a drastic and unpredictable change in the resulting hash. We used our hash generation algorithm to compute the avalanche effect following the steps given below. 

\begin{itemize}
    \item[1.] Select an input message randomly and generate the corresponding hash value ($H_{org}$) using fixed circuit parameters (fixed for one iteration).  
    \item[2.] Randomly Change one bit of the message and generate the corresponding hash value ($H_{mod}$) using the same circuit parameters as step 1. 
    \item[3.] Calculate the Hamming distance between the hash values of the original and modified messages.
    \item[4.] Computing the percentage change, i.e., the avalanche between the hash values of the original and modified messages as,  
    \begin{align}
        &\nonumber\text{Avalanche} \\ &=\frac{\text{Hamming Distance($H_{org}$, $H_{mod}$)}}{\text{Length of the Hash}} \times 100.
    \end{align}
    \item[5.] Repeat steps 1 to 4 $1000$ times with different circuit parameters and compute the average of the avalanche. 
\end{itemize}

\begin{table}[htbp!]
    \centering
    \begin{tabular}{cccccc}
    \hline\\
 \textbf{Unitary Type} & \textbf{Matrix Size} & \textbf{Mean Av.} & \textbf{SEM} & \textbf{Max Av.} \\
 \hline\\
        CUE & $2 \times 2$ & $49.24$\% &$0.28$ & $92.50$\%\\
        CUE & $4 \times 4$ & $51.94$\% & $0.16$ & $80.00$\%\\
        CUE & $8 \times 8$ & $52.69$\% & $0.20$ & $72.50$\%\\
        CUE & $16 \times 16$ & $51.80$\% & $0.17$ & $67.50$\%\\
        CUE & $32 \times 32$ & $50.75$\% & $0.18$ & $61.25$\%\\
\hline\\
        COE & $2 \times 2$ & $49.23$\% & $0.28$ & $95.00$\%\\
        COE & $4 \times 4$ & $52.03$\% & $0.24$ & $72.50$\%\\
        COE & $8 \times 8$ & $52.30$\% & $0.20$ & $76.25$\%\\
        COE & $16 \times 16$ & $51.84$\% & $0.16$ & $65.00$\%\\
        COE & $32 \times 32$ & $50.44$\% & $0.19 $ & $62.50$\%\\
\hline\\
    \end{tabular}
    \caption{Avalanche effect Analysis for matrices from Circular Unitary Ensemble and Circular Orthogonal Ensemble of different dimensions. SEM refers to the standard error of mean.}
    \label{AVA_160}
\end{table}

\begin{figure}
    \includegraphics[scale=0.25]{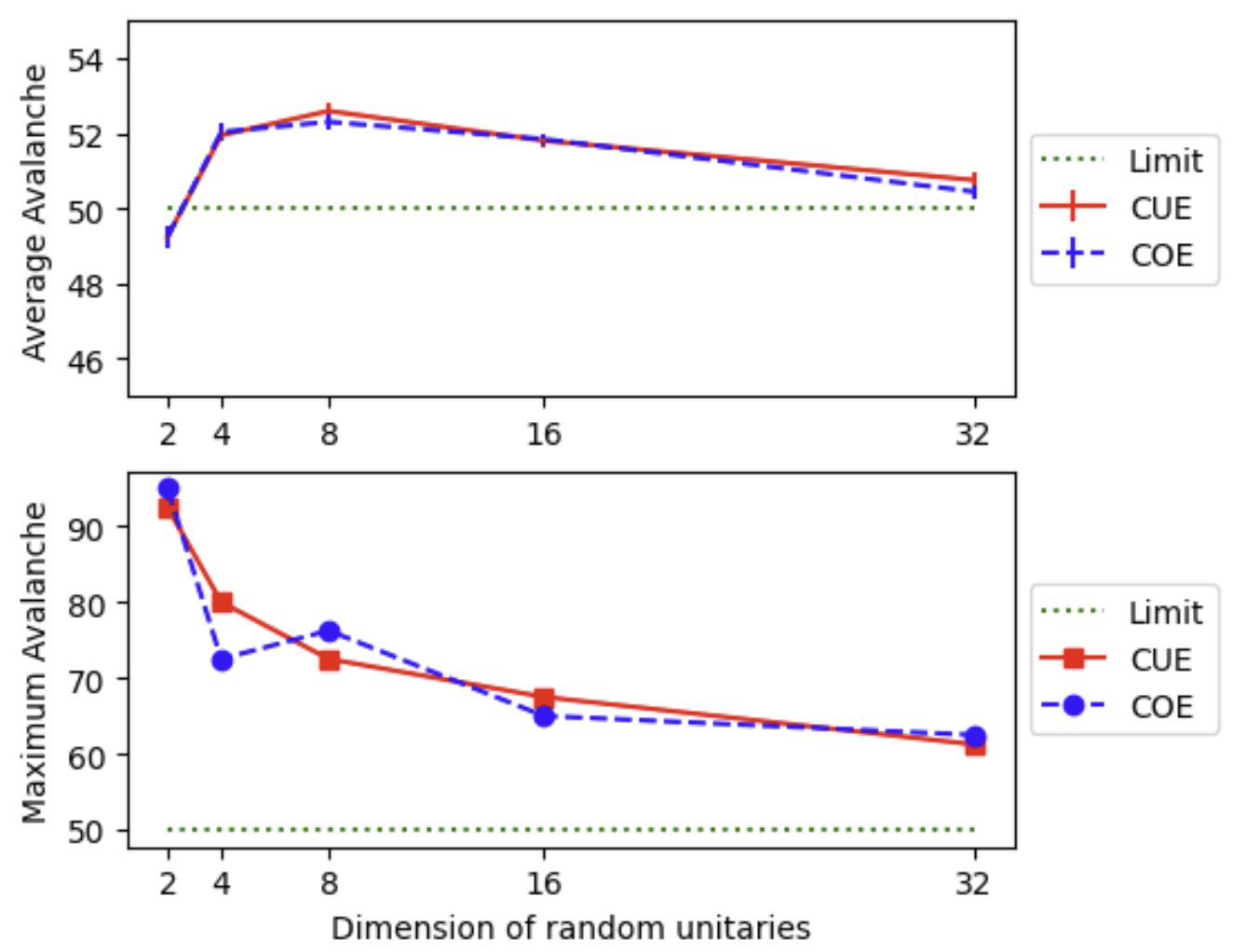}
    \caption{Avalanche effect Analysis for matrices from Circular Unitary Ensemble and Circular Orthogonal Ensemble of different dimensions. The SAC limit is also shown in the diagram.}
    \label{A160}
\end{figure}

We have computed the average avalanche over $1000$ repetitions for different dimensions of the random unitary matrices. Consequently. We have considered two distinct ensembles of random matrices (CUE and COE), to investigate the size and type of random unitary matrix that provides a better avalanche effect. For this analysis, we have kept the size of the position resister fixed at $5$ qubits. To compute one average avalanche, we have selected $1000$ random matrices from each of ensembles, each time randomly sampling an $8$-bit message, as well as taking a random combination of target qubits on the position registers. Our results are provided in Table~\ref{AVA_160}, as well as in Fig.~\ref{A160}. As shown in the table \ref{AVA_160} with the increment of the unitary size matrix the mean avalanche attains a better value, before dropping again. However, for both the cases, for unitaries with dimension greater than or equal to $4 \times 4$, the mean of the avalanche is above the Strict Avalanche Criterion (SAC) limit ($50\%$), where in all the cases, the maximum avalanche found is above SAC limit. The maximum avalanche is shown to decrease with the increasing dimension of unitaries.  

%% REASONS ##

\subsubsection{Reliability}
A reliable hash function must always produce the same value for a given input. This property ensures predictability and consistency.

\begin{itemize}
    \item[1.] Select a message randomly and generate the corresponding hash value.
    \item[2.] For $n$ iterations, generate the hash value for that particular message.
    \item[3.] Repeat steps (1) to (2) $N$ times.
\end{itemize}
We performed the above steps with $N=1000$ times for $1000$ different initial messages of varying length and got reliability as $1$ during simulation of our algorithm with qiskit statevector simulator. This ideal outcome is due to reversible, unitary nature of quantum computation. We understand that this is an idealistic settings, and in a real quantum device with measurements, the reliability will differ from $1$. However, these scenarios are beyond the scope of this work, which presents only theoretical scheme. We plan to report on the experimental outcomes on quantum devices at a future work. 
\subsubsection{ Resistance to Birthday Attacks }
Birthday attacks are a class of cryptographic attacks that exploit the birthday paradox. The birthday paradox states that in a set of randomly chosen people, there is a surprisingly high probability that at least two people share the same birthday. It is also known as 'brute-force' attack, where the mathematics beyond the birthday paradox is exploited to establish that it is possible to find a collision of a $M-$bit hash function with $50$\% chance in $O({\sqrt{2^M}})$ tries.
The hash length of 160 bits means that the output of the hash function is a string of 160 binary digits (0s and 1s). This length is significant in this context, as under birthday attack, an attacker will have to try at least $2^{80}$ times before finding a collision with $50$\% chance, i.e., a longer hash length increases the difficulty of finding collisions through a birthday attack. Since our protocol presents a way to compute a large Hash function with significantly lower resource cost, one can infer that it shows strong resistance towards birthday attack. 

\subsection{384-bit Hash Generation and Analysis:}\label{H384}
We now present our results for the 384-bit Hash generation.

\subsubsection{Sensitivity of Hash value to Message}
We have analysed the sensitivity of the generated $384$-bit Hash, considering same conditions as provided in Section~\ref{RnD}, subsection~\ref{H160}. We show that the presented hash function is highly sensitive to the conditions considering input message $'01100010110101001'$, and randomly choosing a matrix of dimension $2 \times 2$ from CUE and COE ensembles. The generated hash values are given in Tables~\ref{Sens_CUE_384}, \ref{Sens_COE_384} respectively in additional information section, section~\ref{appen}, due to their larger bit-length. The number of qubits in position register here is $5$, message register has $2$ qubits, coin register has a single qubit, and the ancila register has $6$ qubits.  

\subsubsection{Collision Resistance}
\begin{table}[htbp]
    \centering
    \begin{tabular}{ccccc}
    \hline \\
        \textbf{Matrix}& \textbf{Dimension} & \textbf{Qpos} & \textbf{Total Collision} & \textbf{Collision Rate} \\
    \hline \\
        CUE & $2 \times 2$ & 5 & 0 & 0\\
        CUE &$2 \times 2$ & 6 & 0 & 0 \\
        CUE &$2 \times 2$ & 7 & 0 & 0 \\
        CUE &$4 \times 4$ & 5 & 0 & 0 \\
        CUE &$4 \times 4$ & 6 & 0 & 0 \\
        CUE &$4 \times 4$ & 7 & 0 & 0 \\
        CUE &$8 \times 8$ & 5 & 0 & 0 \\
        CUE &$8 \times 8$ & 6 & 0 & 0 \\
        CUE &$8 \times 8$ & 7 & 0 & 0 \\
        \hline \\
        COE & $2 \times 2$ & 5 & 0 & 0\\
        COE & $2 \times 2$ & 6 & 0 & 0 \\
        COE & $2 \times 2$ & 7 & 0 & 0 \\
        COE & $4 \times 4$ & 5 & 0 & 0 \\
        COE & $4 \times 4$ & 6 & 0 & 0 \\
        COE & $4 \times 4$ & 7 & 0 & 0 \\
        COE & $8 \times 8$ & 5 & 0 & 0 \\
        COE & $8 \times 8$ & 6 & 0 & 0 \\
        COE & $8 \times 8$ & 7 & 0 & 0 \\
        \hline
        \end{tabular}
        \caption{Collision Analysis table for  of Different Input types. 'Qpos' refers to the number of qubits in the position register. Total number of times Hash is generated for each individual case (each row of the table) is $1000$.}
        \label{collision384}
\end{table}
Similar to the case with $160$-bit Hash generation, We perform the collision resistance test considering different aspects of our Hash generation protocol, and tabulate them in Table~\ref{collision384}. As can be seen, in this case too, we found zero collision.

\subsubsection{Avalanche effect}
\begin{table}[htbp!]
    \begin{tabular}{cccccc}
    \hline\\
 \textbf{Unitary Type} & \textbf{Matrix Size} & \textbf{Mean Av.} & \textbf{SEM} & \textbf{Max Av.} \\
 \hline\\
        CUE & $2 \times 2$ & $47.98$\% &$0.21$ & $76.04$\%\\
        CUE & $4 \times 4$ & $50.97$\% & $0.19$ & $67.17$\%\\
        CUE & $8 \times 8$ & $52.22$\% & $0.18$ & $71.35$\%\\
        CUE & $16 \times 16$ & $51.76$\% & $0.15$ & $69.79$\%\\
        CUE & $32 \times 32$ & $50.49$\% & $0.20$ & $59.89$\%\\
\hline\\
        COE & $2 \times 2$ & $48.53$\% & $0.26$ & $83.33$\%\\
        COE & $4 \times 4$ & $51.63$\% & $0.20$ & $68.23$\%\\
        COE & $8 \times 8$ & $51.97$\% & $0.23$ & $69.27$\%\\
        COE & $16 \times 16$ & $51.52$\% & $0.15$ & $66.67$\%\\
        COE & $32 \times 32$ & $50.65$\% & $0.21 $ & $58.85$\%\\
\hline\\
    \end{tabular}
    \caption{Avalanche effect Analysis for matrices from Circular Unitary Ensemble and Circular Orthogonal Ensemble of different dimensions for Hash value with bit-length $384$. SEM refers to the standard error of mean.}
    \label{AVA_384}
\end{table}

\begin{figure}
    \includegraphics[scale=0.59]{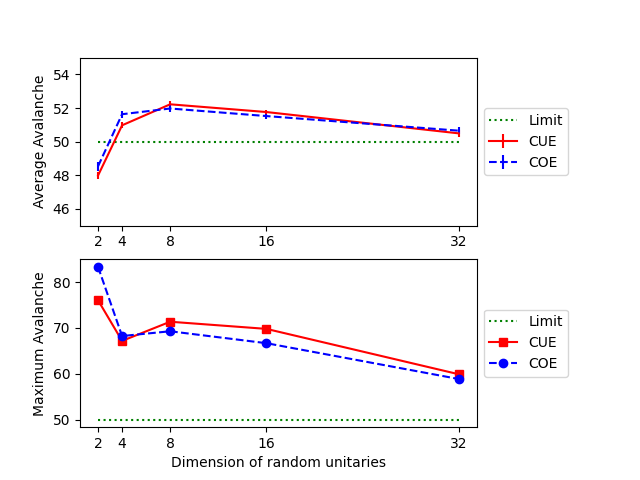}
    \caption{Avalanche effect Analysis for matrices from Circular Unitary Ensemble and Circular Orthogonal Ensemble of different dimensions for generated hash value with bit-length $384$. The SAC limit ($50\%$) is shown in the diagram.}
    \label{AVA_384_g}
\end{figure}

We have performed the avalanche analysis considering $1000$ different input messages taking two different types of random unitary matrices with varying dimension, similar to the $160-$bit Hash generation. Since we have kept the number of qubits in the position register fixed at $5$, the maximum allowed dimension of the unitary can be $32$. 
As shown in Table~\ref{AVA_384} and Fig.~\ref{AVA_384_g}, the average avalanche of the Hash function is above the SAC limit for all the considered dimension of the random unitaries other than $2 \times 2$. Also, the maximum avalanche stays well above the SAC limit. 

\subsubsection{Reliability}
We performed reliability analysis as mentioned in the Subsection~\ref{H160} and got reliability as $1$ in this case as well. 

\subsubsection{Resistance to Birthday Attacks} For this Hash value, the brute force attacker needs to try at least $O(2^{172})$ times before finding a collision with $50\%$ chance, making it even harder to find a collision. 

\subsection{Comparison of FQHF with Quantum walk-based hash generation schemes:}
The major difference between FQHF and other quantum walk based protocols is that in this protocol, we have eradicated the need for classical post-post processing algorithms, and created a prepare and measure like protocol for Hash generation. We consider other metrics for Hash values, and provide a comparison between different Hash generation schemes. 
\subsubsection{Better Collision Resistance}
In FQHF, we get $0$ collision rate under idealistic assumptions. However, QWH algorithms that uses classical post processing reports a collision rate. We tabulate them in Table~\ref{Tab_CRC}.
\begin{table}[!htbp]
    \begin{tabular}{cccccc}
       
         \textbf{Ref. \cite{b24}} & \textbf{Ref. \cite{b36}} & \textbf{Ref. \cite{b37}} & \textbf{Ref. \cite{b38}} & \textbf{Ref. \cite{b39}} &
         \textbf{FQHF}\\
        
        23.12 & 6.33 & 9.32 & 1.16 & 1.95 & 0.0\\
      
    \end{tabular}
\caption{Collision rates for different Schemes.}
\label{Tab_CRC}
\end{table}
\subsubsection{Better Speed Performance}
In our method, $O(L)$ measurements are required to obtain the complete probability distribution at the ancilla register, to obtain a hash value of bit length of the order $L$. This is true for QWH+Classical Post-processing algorithms as well. Also, both types of algorithms uses similar number of qubits, as well as operational resources. However, our approach avoids using traditional computations after the initial quantum hashing process, making it intrinsically faster than the other QWH algorithms that rely on classical post-processing. This improve the overall speed and efficiency of the hashing process. This makes our method a strong candidate for practical applications where speed is crucial.

\subsubsection{Fully quantum process}
FQHF generally offer more robust security guarantees due to their exclusive reliance on quantum operations. The algorithm is intrinsically less vulnerable to classical attacks targeting the post-processing steps. Compared to other QWH algorithms, an extra layer of randomness is added to the Hash generation scheme in terms of the DQC1 operations. However, the security of FQHF relies on the robustness of quantum algorithms and the absence of efficient quantum attacks.

\section{Conclusion}\label{conc}
Our research introduces a Fully Quantum Hash Function (FQHF) protocol designed within the quantum walk framework. The FQHF uses polynomial operational resources and logarithmic spatial resources. It has measurement complexity linear in bit-length of the Hash function. No complex post-processing operation is needed for the protocol. Further, we have shown that our hash function exhibit good sensitivity to the message, $0$ collision rate, reliability $1$, and average avalanche over SAC limit under idealistic (noiseless) quantum computation scenarios. The large bit-length of output Hash ensures resistance to birthday attacks as well. Further, We have provided a comparison of our protocol with varying system parameters considering collision, reliability, and avalanche. We found the reliability and collision to be system independent, whereas, avalanche shows dependence on the size of the random unitary matrices used. In both the example hash generation cases, we have found maximum avalanche decreases with increase in unitary dimension, and average avalanche increases in the beginning, and then decreases slightly. This is anticipated, as with very low random unitary dimension, the ancilla qubit in DQC1 scheme only encapsulates 
a portion of randomness present in the circuit, where as with very high unitary dimension, ($\approx$ number of qubits in position register), the randomness in the scheme decreases, since all DQC1 blocks use the same target ancillas. It will be of  interest to study and quantify this correlation in detail in future. However, we did not find any substantial correlation between type of random unitary matrices used and performance of the hash function.   

Unlike classical hash functions that are vulnerable to quantum decryption, FQHF leverages quantum mechanical principles, offering unparalleled data integrity and security. The zero collision rate of FQHF indicates its robustness against cryptographic attacks. Our approach emphasizes the importance of fully quantum processes, enhancing security by avoiding classical post-calculation steps. Further studies involving different quantum walk schemes and unitary matrices will be useful to understand the full potential of this work. Also, it will be interesting to see how this protocol works under usual noises present in a quantum computer.  

\bibliographystyle{unsrt}
%\bibliography{ref}

%\noindent LaTeX formats citations and references automatically using the bibliography records in your .bib file, which you can edit via the project menu. Use the site command for an inline citation, e.g., \cite{Hao:gidmaps:2014}.

%For data citations of datasets uploaded to, e.g., \emph{figshare}, please use the \verb|howpublished| option in the bib entry to specify the platform and the link, as in the \verb|Hao:gidmaps:2014| example in the sample bibliography file.

%\section{Acknowledgements}
%\section{Author contributions statement}
%A.A. conceived the experiment(s),  A.A. and B.A. conducted the experiment(s), C.A. and D.A. analyzed the results.  All authors reviewed the manuscript. 

\onecolumn

\section{Additional information}\label{appen}

We present the results for sensitivity analysis of the Hash function of bit-length $384$ in Tables \ref{Sens_CUE_384}, and \ref{Sens_COE_384}.  

\begin{table}[!hbt]
    \begin{tabular}{c|c}
    \hline \\
        \textbf{Condition} & \textbf{Generated Hash when CUE is used} \\
    \hline \\
        1 &  '20A61A1470809D59645E6490DCB275D06D846D958EA4C52B7B973B34F0C175F4D1AA8EADE2FF7FBAC8E183E8BCCF1C32'\\
    \hline \\
        2 & '20A9186A6D3638C18409E01C5944902C99E534F6D9DF57457C75570C18224D1C32BEDFFDB2EAE9F3EEE3E61CF1A2AE3A'
    ' \\
    \hline \\
        3 & '14720A5579E561ADE9D6B080B6FE7B490F7F1842C95A45266D9D36A83A01E93E11BAC34FFBC38C81D89F79CC328E1CF1' 
     \\
     \hline \\
        4 & '30E71E9A43E5367DB47F577710620A51661A0804902C91C56D95D50C14D18A0863CB0C73BACB6FFBCF7FAA8A6BEB8E7B' \\
    \hline \\
         5 &'EF9698AE9288594E1243ACC61317A809C02A8D975B85FC3238C24D54B3E05E2FFD147BED4D10C1FBCBACDF5D369E5926'\\
    \hline 
    \end{tabular}
    \caption{Output hash in Hexadecimal format for input message $'01100010110101001'$ under different initial conditions, generated with  $5$ position qubits, $1$ coin qubits, and $6$ ancilla qubits. The controlled unitary matrices after quantum walk are $5$ random $2 \times 2$ matrices from Circular Unitary Ensemble with a control on each ancilla qubit, applied to random qubits on the position register.}
    \label{Sens_CUE_384}
\end{table}

\begin{table}[!hbt]
    \begin{tabular}{c|c}
    \hline \\
        \textbf{Condition} & \textbf{Generated Hash when COE is used}  \\
    \hline \\
        1 & 'C10804D1499EB8C72CF0EF82692A88A182390B9C518868B2545D559DE9B7BCD76DF4FFC36D3AC9A5927BEC78F35E7D'\\
    \hline \\
        2 & '300204364049805B2896172D418A6FACF2D4774459C0ABAA3959C70E3F38D719821A2F79FFB7DB6BEE9EDD74F3D925B2' \\ 
    \hline \\
        3 & '9EB8C0107BC33082E6A8F881146B8234928EDE0EF3B10914A17FD37186F687A7F2B51955A56F9275969EC3CD38C7DD79'\\
    \hline \\
        4 & '30800D2440609231672CF865728B188A6AEF38A41DA6D6541EA0EE470D02CD51B76DFC75E3CCB6EFF79AE7D5FA4FE496'\\
    \hline \\
        5 & 'FD073B518F9177A559CF76DFCB6BFC00CAF811A2307B4B93E415CDF6A149C758E74962CF8A6B282A03A40E91ED865086'
\\
    \hline
    \end{tabular}
    \caption{Output hash in Hexadecimal format for input message $'01100010110101001'$ for different conditions, generated with 5 position qubits, 1 coin qubits, and 5 ancilla qubits. The controlled unitary matrices after the quantum walk are $5$ random $2 \times 2$ matrices from Circular Orthogonal Ensemble with a control on each ancilla qubit, applied to random qubits on the position register.}
    \label{Sens_COE_384}
\end{table}
\twocolumn

\end{document}